# A discrete time evolution model for fracture networks

Gábor Domokos[1,2,3] and Krisztina Regős[1,2]

[1] Department of Morphology and Geometric Modeling, Budapest University of Technology and Economics, H-1111 Budapest, Műegyetem rkp 3., K220.
[2] MTA-BME Morphodynamics Research Group, Budapest University of Technology and Economics, H-1111 Budapest, Műegyetem rkp 3., K220.
3 Corresponding author: domokos@iit.bme.hu

Abstract

We examine geophysical crack patterns using the mean field theory of convex mosaics. We assign the pair $(\bar{n}^*, \bar{v}^*)$ of *average corner degrees* to each crack pattern and we define two local, random evolutionary steps $R_0$ and $R_1$, corresponding to secondary fracture and rearrangement of cracks, respectively. Random sequences of these steps result in trajectories on the $(\bar{n}^*, \bar{v}^*)$ plane. We prove the existence of limit points for several types of trajectories. Also, we prove that *cell density* $\bar{\rho} = \frac{\bar{v}^*}{\bar{n}^*}$ increases monotonically under any admissible trajectory.

Keywords

Fracture network, evolution model, discrete dynamical system, evolution model, tessellation

# 1. <u>Introduction</u>

Fragmentation is one of the most ubiquitous natural processes and the efforts to decode its geometry have been at the forefront of geophysical research [1-8]. While many aspects of fragmentation are inherently three dimensional, the 2D aspects of the phenomenon are interesting in their own right: the most visible fingerprints of fragmentation are surface fracture patterns (also called 2D fracture networks) on various scales [4], ranging from mud cracks resulting from desiccation [6,7,8] (see Figure 1 for two examples) through basalt columns [3] to the pattern defined by the tectonic plates [2,9]. In [2] the geometric theory of tilings [13,14], in particular, the mean field theory of convex mosaics [10,12] was applied to identify fracture networks with a point on the so-called symbolic plane, spanned by two characteristic geometric features, the *nodal and cell degrees* [10,14] (see Definitions 4 and 5 in Section 2).

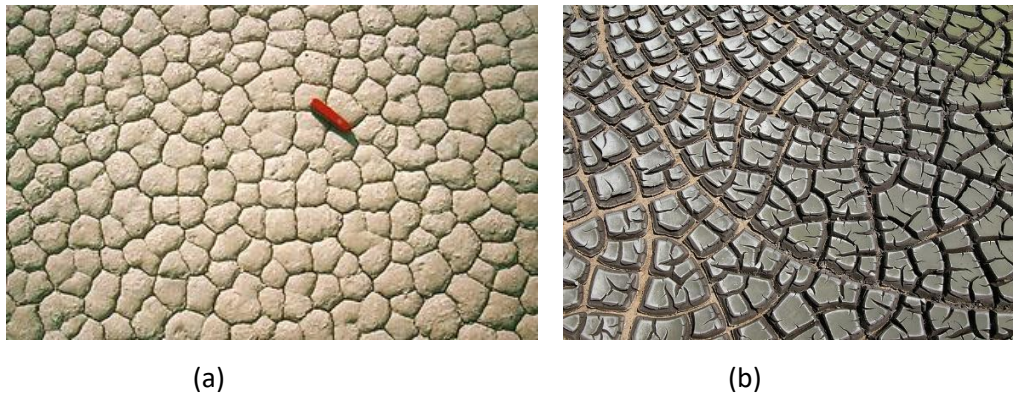

(a)  (b)

**Figure 1:** Desiccation crack patterns in mud, also discussed in [2]. Observe different geometric tiling patterns. Photo credit: (a): Charles E. Jones University of Pittsburgh, Pittsburgh, PA (b):Hannes Grobe, Alfred Wegener Institute, Bremerhaven, Germany

However, fracture networks are not static objects; they *evolve* in various manners. Evolution models are popular mathematical tools, used also in operations research [16]. This particular evolution may be modeled by regarding an initial fracture network (to which we will refer as *primary network*) and then consider a family of discrete, local events under the random sequence of which the primary network evolves. While these events have been described in [2], the evolution model has not been developed and this is the goal of the current paper. Such a model would, for example, admit the following

**Question 1:** *Could the pattern in Figure 1(a) have evolved from the pattern in Figure 1(b) or vice versa?*

After constructing our model in Sections 2 and 3, we will address Question 1 in Section 4, showing that the (a)→(b) evolution is not admissible in the model but the (b) →(a) is.

In the model we will consider two types of local events which evolve the fracture network. One such local event is undoubtedly *secondary fracture* where an existing fragment particle (produced in primary fracture) is being split into two parts along a (random) fracture line; this can be observed on rock outcrops. Another local event is when, in the process of crack healing and rearrangement "T" nodes evolve into "Y" nodes; this can be observed in drying mud and also in columnar joints [6,8].

Our evolution model will only consider these two steps. Needless to say, the model could be refined by considering other local events. However, even this simple model suffices to show the principles of how such models operate in general and, as we will show, the model based on these two steps has already rich dynamics and promises to explain several features of the observed natural processes.

Our paper is structured as follows: in Section 2 we introduce all necessary mathematical concepts. Section 3 is dedicated to the development of the model and the main results and in Section 4 we draw conclusions.

## 2. **Mathematical concepts**

As models of 2D fracture networks we consider *convex, normal tilings* which we define below, following [11,12]:

**Definition 1.**
*Normal tilings are 2 dimensional tessellations where each cell is a topological disk, the intersection of each of the two cells is either a connected set, or the empty set and the cells are uniformly bounded from below and above.*

**Definition 2.**

*Convex tilings are 2 dimensional tessellations where each cell is convex.*

**Remark 1.**

As a direct consequence of Definitions 1 and 2, we see immediately that all cells of normal, convex tilings of the Euclidean plane are finite, convex polygons (see also [11].) We can also see that normal tilings of infinte domains (such as the Euclidean plane) have infinitely many cells while normal tilings of finite (compact) domains have a finite number of cells.

**Definition 3.**

*A node of a convex, normal tiling is a point where at least 3 cells overlap.*

**Remark 2.**

As a consequence of Remark 1 and Definition 3, we can see that in a convex, normal tiling the boundary of each cell contains at least 3 nodes.

**Definition 4.**

*In a normal, convex tiling the combinatorial degree v of a cell is equal the number of nodes on its boundary and the combinatorial degree n of a node is equal to the number of cells overlapping at that node.*

Following [15], we also introduce a concept, which is more sensitive to the actual shape of the cells:

**Definition 5.**

*In a normal, convex tiling the corner degree v\* of a cell is equal to the number of its vertices and the corner degree n\* of a node is equal to the number of vertices overlapping at that node.*

**Remark 3.**

We can immediately see that for all cells and nodes we have $v^* \leq v, n^* \leq n$.

**Definition 6.**

We call a node regular if $n^* = n$. We call a normal, convex tiling regular is all nodes are regular. We denote the number of all nodes by V and the number of irregular nodes by $V_I$ and the quantity $r = \frac{V - V_I^*}{V}$ is called the regularity of the tiling, $r = 1$ corresponding to regular tilings. (We remark that for regular tilings we have $v^* = v$ for all cells. We also see that for regular nodes we have $n^* \geq 3$ whereas for irregular nodes we have $n^* \geq 2$.)

The quantities $v, v^*n, n^*$ can be easily averaged on finite mosaics:

**Definition 7.**

Let us denote the number of faces (cells), edges and vertices (nodes) of a finite portion of a convex, balanced tiling by F,E,V, respectively. We call the quantities

(1) $$\bar{n} = \frac{N}{V}$$

(2) $$\bar{v} = \frac{N}{F}$$

the combinatorial degrees of the tiling, where $N = \sum_{i=1}^{F} v_i = \sum_{i=1}^{V} n_i$ and $n_i, v_i$ denote the combinatorial degrees of the $i^{th}$ node and cell, respectively. Similarly, the quantities

(3) $$\bar{n}^* = \frac{N^*}{V}$$

(4) $$\bar{v}^* = \frac{N^*}{F}$$

are called the corner degrees of the tiling, where $N^* = \sum_{i=1}^{F} v_i^* = \sum_{i=1}^{V} n_i^*$. We will refer to $F, V, N, N^*$ as the fundamental quantities of the tiling.

The average degrees can also be defined for infinite normal tilings:

**Definition 8.**

Select a sphere $B(X, \rho)$ with center X and radius $\rho$, and calculate the averages $\bar{n}(X, \rho), \bar{v}(X, \rho)$ and $\bar{n}^*(X, \rho), \bar{v}^*(X, \rho)$. The radius $\rho$ is then increased to infinity. We say that the averages $\bar{n}, \bar{v}, \bar{n}^*$ and $\bar{v}^*$ of the combinatorial and corner degrees exist if the limits $lim_{\rho \to \infty} \bar{n}(X, \rho)$, $lim_{\rho \to \infty} \bar{v}(X, \rho)$, $lim_{\rho \to \infty} \bar{n}^*(X, \rho)$ and

$lim_{\rho\to\infty}\bar{v}^*(X,\rho)$ exist and they are independent from X. Tilings with property are called *balanced* [8] tilings.

## Remark 4.

Henceforth we only consider normal, convex tilings and for all infinite tilings we require that they are balanced.

## Definition 9.

The $[\bar{n}, \bar{v}]$ plane is called the *(combinatorial) symbolic plane* and the $[\bar{n}^*, \bar{v}^*]$ plane is called the *(metric) symbolic plane*.

## Remark 5.

Based on [11,12], for infinite, balanced, convex (a) regular and (b) irregular tilings of the Euclidean plane we have:

(5) $\quad\quad\quad\quad (a) \quad \bar{v} = \dfrac{2\bar{n}}{\bar{n}-2} \quad\quad (b) \quad \bar{v} = \dfrac{2\bar{n}}{\bar{n}-r-1}$

and there exist no infinite tilings for $\bar{v} > \dfrac{2\bar{n}}{\bar{n}-2}$ and no convex tilings for $\bar{v} > 2\bar{n}$.

## Definition 10.

Let *M* be a finite part of a regular, normal, convex tiling and let *M* have F faces (cells) and V vertices (nodes). Then we call $\bar{\rho}(M) = \dfrac{V}{F}$ the *cell density* of *M*. For infinite tilings *M* we use the limit process described in Definition 6.

## Remark 6.

We can immediately see that $\bar{\rho}(M) = \dfrac{\bar{v}(M)}{\bar{n}(M)} = \dfrac{\bar{v}^*(M)}{\bar{n}^*(M)}$. So, the cell density can be expressed by the average cell degree and average nodal degree, but its value is *independent of the definition of these degrees*. We can also see that $\bar{\rho}(M) = constant$ lines correspond to rays passing through the origin of the symbolic plane.

# 3. An evolution model for fracture networks

Our model is inspired by [2], where planar fracture patterns have been analyzed and classified in the symbolic plane, based on their nodal and cell averages. Here we take one further step and build a model to study not only their current state but also their evolution in the symbolic plane.

We start by defining our main hypothesis. While we believe that this hypothesis captures several aspects of physical fragmentation, our goal is not just to define this particular evolution model but also to demonstrate how such a model is constructed and how it operates.

### Hypothesis 1.
A. The fracture network is a finite domain of a convex, balanced tiling of the Euclidean plane.

B. Evolution of the crack pattern takes place in a series of discrete events, consisting of 2 step-types ($R_0, R_1$) to be detailed below. The order of the step-types is important: we assume that the first $k$ pieces of $R_1$ steps are preceded by at least ($k/2$) pieces of $R_0$ steps. (This is explained in detail below when we define step $R_1$.)

C. (1) During *secondary cracking*, one cell of the primary crack network is split into two parts along a straight line segment, connecting two points belonging to the relative interior of two different edges of the cell. ($R_0$ type step). *It is easy to see that $R_0$ type steps retain the convexity of the initial mosaic.* See Figure 2.

(2) During *crack healing-rearrangement*, the edges and nodes of the crack network are rearranged so that "T" nodes evolve into "Y" nodes [4] and each such event corresponds to an $R_1$-type step. Obviously, this step can only be performed on an irregular "T" node. Since we assumed the initial mosaic to be regular and one step of type $R_0$ generates two irregular nodes, the first $k$ pieces of $R_1$ steps must be preceded by at least ($k/2$) pieces of $R_0$ steps. See Figure 2.

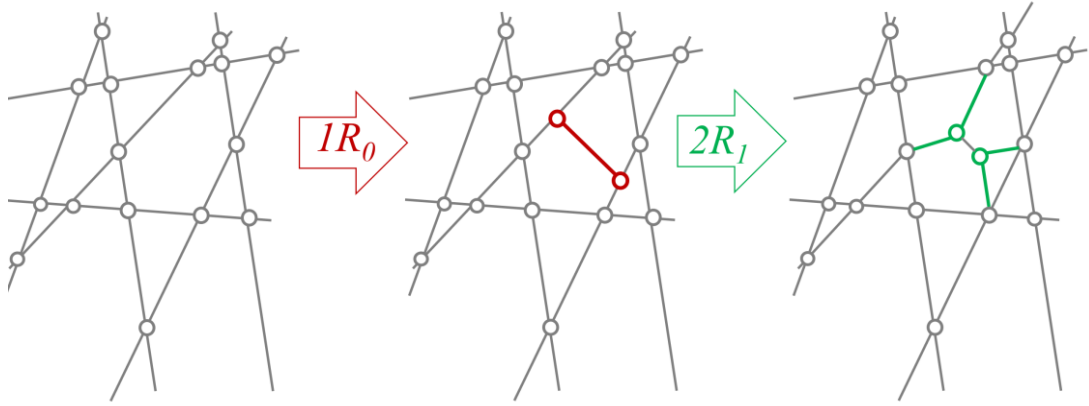

**Figure 2:** Geometry of evolution steps $R_0$ and $R_1$ used in the model based on Hypothesis 1. Observe that the first $k$ pieces of $R_1$ steps must be preceded by at least $(k/2)$ pieces of $R_0$ steps.

For clarity, below we summarize in Table 1 how the steps $R_0$ and $R_1$ operate on the fundamental quantities $F, V, N, N^*$ of the tiling. See also Figure 2 for illustration.

| Step type | Name of the step | $N_k$ | $N_k^*$ | $F_k$ | $V_k$ |
|---|---|---|---|---|---|
| $R_0$ | Secondary cracks | $N_{k-1} + 6$ | $N_{k-1}^* + 4$ | $F_{k-1} + 1$ | $V_{k-1} + 2$ |
| $R_1$ | Crack healing rearrangement | $N_{k-1}$ | $N_{k-1}^* + 1$ | $F_{k-1}$ | $V_{k-1}$ |

**Table 1:** Evolutionary equations for the steps $R_0$ and $R_1$ used in the model based on Hypothesis 1.

**Definition 11:** Let $M$ be a finite, convex, normal mosaic characterized by an initial state $(\bar{n}_I^*, \bar{v}_I^*)$ and let us denote by $(\bar{n}_p^*, \bar{v}_p^*)$ the limit point on the symbolic plane, reached after applying an infinite random sequence of $R_0$ and $R_1$ steps, with respective probabilities *1-p, p* and we call this a *p-trajectory*.

**Remark 7:**

We mention two special *p*-trajectories: *0*-trajectories correspond to sequences consisting entirely of $R_0$-type steps and *1*-trajectories correspond to sequences consisting entirely of $R_1$-type steps. We also mention that if the mosaics corresponding to any finite number of evolution steps $R_0$ and $R_1$ are, like the initial mosaic $M$, finite, convex, normal tilings. However, the limit point $(\bar{n}_p^*, \bar{v}_p^*)$ does not correspond to a normal mosaic, it should be regarded as a point of the symbolic plane.

The following lemma applies to general $p$-trajectories:

**Lemma 1:**

*Let $M$ be a finite, normal, convex mosaic characterized by an initial state $(\bar{n}_I^*, \bar{v}_I^*)$. Then, for all $p$-trajectories we have $(\bar{n}_p^*, \bar{v}_p^*) = \left(\frac{4-3p}{2-2p}, \frac{4-3p}{1-p}\right).$*

**Proof** :

(a) After the first $c(1-p)$ steps of type $R_0$ we have:

$$N^*(c,p) = N_0^* + 4c(1-p); \quad F(c,p) = F_0 + c(1-p); \quad V(c,p) = V_0 + 2c(1-p).$$

(b) After $cp$ steps of type $R_1$ we have:

$$N^*(c,p) = N_0^* + cp; \quad F(c,p) = F_0; \quad V(c,p) = V_0.$$

Now we can compute limit for the mixed trajectory as $c$ approaches infinity. Using (3) and (4) we can write:

$$\bar{n}_p^* = \lim_{c \to \infty} \bar{n}^*(c,p) = \lim_{c \to \infty} \frac{N_0^* + 4c(1-p) + cp}{V_0 + 2c(1-p)} = \frac{4-3p}{2-2p},$$

$$\bar{v}_p^* = \lim_{c \to \infty} \bar{v}^*(c,p) = \lim_{c \to \infty} \frac{N_0^* + 4c(1-p) + cp}{F_0 + c(1-p)} = \frac{4-3p}{1-p}.$$

Q.e.d.

The result of Lemma 1 is illustrated in Figure 3 where we computed standard $p$-trajectories for $p = 0, \frac{1}{2}, \frac{2}{3}, \frac{3}{4}, 1$.

**Remark 8:**

We can see that all limit points lie on the $\bar{v}^* = 2\bar{n}^*$ line. However, the *physically relevant* portion of the trajectory may terminate earlier. This is easiest understood if we consider the following argument: when constructing physically relevant *p*-trajectories we also have to consider Hypothesis 1, C(2) which implies that the necessary condition for an infinite *p*-trajectory to be physical is $p \leq 2/3$, as at least half as many $R_0$ steps are needed as we have $R_1$ steps. For $p \geq 2/3$, the physical part of the trajectory will terminate as the mosaic becomes regular at (or, for finite mosaics, near) the hyperbola defined in (5)(a). Beyond this point, the trajectory exists only in an algebraic sense (as a sequence of numbers) to which we do not attach any direct geometric interpretation. We also mention that, as $p \to 1$, the tangent of the trajectory at $(\bar{n}_I^*, \bar{v}_I^*)$ will approach the $\bar{\rho} = constant$ line with constant cell density passing through $(\bar{n}_I^*, \bar{v}_I^*)$, characterized by $\bar{v}^* = \bar{n}^* \bar{v}_I^* / \bar{n}_I^*$.

**Remark 9:**

So far we described *p*-trajectories on finite domains and these trajectories appear on the symbolic plane as discrete point sequences. Based on Lemma 1, these sequences have limit points and the location of these points is independent of the initial point $(\bar{n}_I^*, \bar{v}_I^*)$, it just depends on the value of *p*. Now we extend this concept for the case where the initial mosaic is a balanced, infinite tiling *M* with averages $(\bar{n}_I^*(M), \bar{v}_I^*(M))$. We regard the limit process described in Definition 8: each finite value of the radius $\rho$ defines a finite mosaic *M(ρ)* and we have a *p*-trajectory on *M(ρ)* with initial point $(\bar{n}_I^*(\rho), \bar{v}_I^*(\rho))$ and final point $(\bar{n}_p^*, \bar{v}_p^*)$, the latter being independent of $\rho$. Let us now regard a sequence of $\rho$ approaching infinity. Then we have $lim_{\rho \to \infty}(\bar{n}_I^*(\rho), \bar{v}_I^*(\rho)) = (\bar{n}_I^*(M), \bar{v}_I^*(M))$. We can run the same limit process after having executed a finite number of steps of a *p*-trajectory and after any finite number of steps we get convergence. So we can see that this process defines a sequence of trajectories which converge, We call the *limiting p-trajectory* with initial point $(\bar{n}_I^*(M), \bar{v}_I^*(M))$ and final point $(\bar{n}_p^*, \bar{v}_p^*)$ the *p*-trajectory associated with the infinite, balanced tiling *M*. We also note that trajectories corresponding to infinite tilings are represented in the symbolic plane by continuous line.

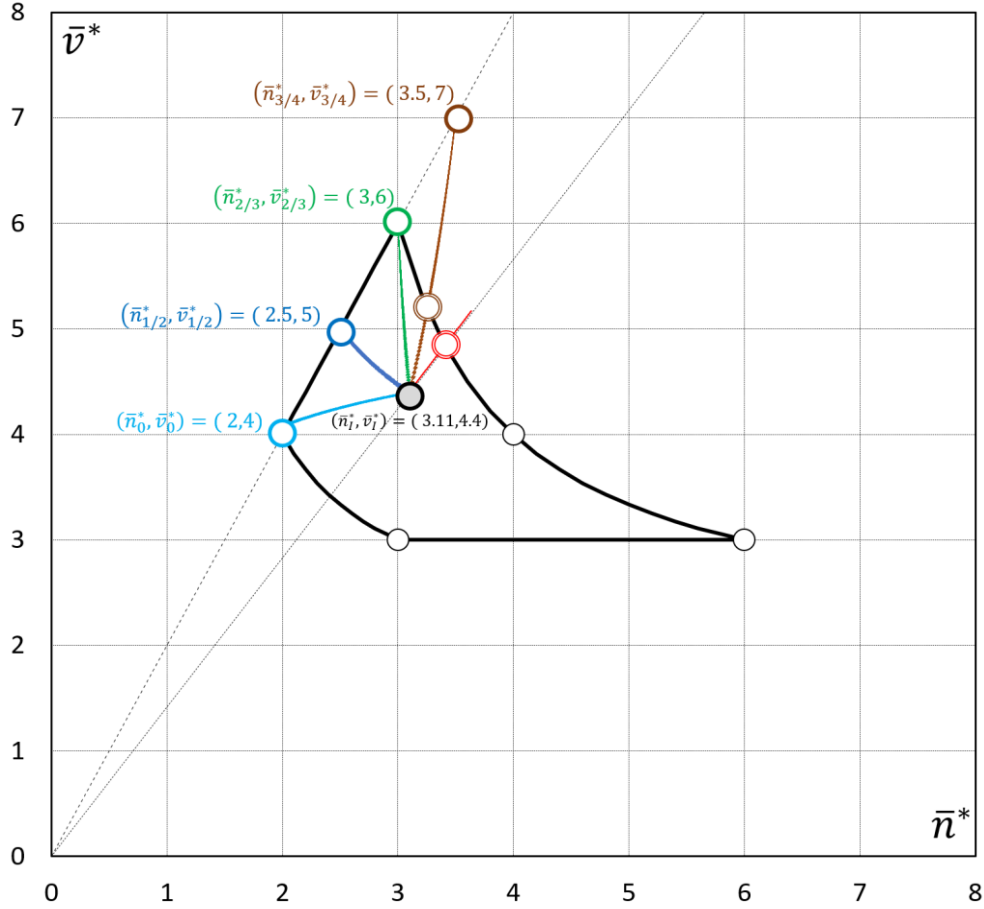

**Figure 3:** Numerical simulation of *p*-trajectories on the symbolic plane. With initial mosaic at $(\bar{n}_I^*, \bar{v}_I^*) =$ (3.11,4.4) (black circle with grey fill) we show 5 trajectories corresponding to p =0, 1/2, 2/3,3/4, 1, respectively. Limit points marked by colored circles with white interior. As predicted by Lemma 1, all limit points are on the $\bar{v}^* = 2\bar{n}^*$ line. Domain occupied by convex mosaics shown with black line, cf [11]. Terminal points of physical trajectories marked by colored circles with double line. Dashed black lines correspond to $\bar{v}^* = 2n^*$ and $\bar{v}^* = \bar{n}^* \bar{v}_I^* / \bar{n}_I^*$, respectively

Needless to say, $p$-trajectories represent only a rather restricted class of geophysical evolution processes for fracture networks. In such a process, the ratio of $R_0$-type and $R_1$-type evolution steps remains constant. However, our model also admits statements of more general type. Relying on definitions 10 and 11 we make the following observation:

**Lemma 2:**

*In the discrete-time fracture network evolution model formulated in Hypothesis 1, the cell density $\bar{\rho}(M)$ increases monotonically over time.*

**Proof:**

Since there are two steps in the model ($R_0$ and $R_1$), so, if we can show separately for both step-types that the cell density does not decrease, we have confirmed the statement of Lemma 2.

In the case of $R_0$-type steps we have $N^*_{k_0} = N^*_{k_0-1} + 4$; $F_{k_1} = F_{k_1-1} + 1$; $V_{k_1} = V_{k_1-1} + 2$ (see Table 1.) Note that from Definition 10 we have $\bar{\rho}(M) = \frac{V}{F}$, and we also note, based on [11], that for any convex mosaic we have $\bar{\rho}(M) \leq 2$ (see also Remark 5). Based on these observations we see that cell density is increasing strictly monotonically under $R_0$-type steps.

In the case of $R_1$-type steps we have $N^*_{k_1} = N^*_{k_1} + 1$; $F_{k_2} = F_{k_2-1}$; $V_{k_2} = V_{k_2-1}$ (see Table 1), so here the cell density does not change.

As described above, in the crack network evolution model formulated in Hypothesis 1, the cell density is either constant or increasing in each step. Q.e.d.

So far, we did not make any additional restrictions on the initial mosaic $M$ beyond requiring that it should be a convex, normal tiling. However, typical primary fracture networks have special position on the symbolic plane; in [2] it was argued that if the network is created by long straight fracture lines, then this corresponds to the point $(\bar{n}^*, \bar{v}^*) = (4,4)$ of the symbolic plane. Motivated by this geophyiscal observation we formulate

**<u>Lemma 3:</u>**
*In the discrete-time fracture network evolution model formulated in Hypothesis 1, if we choose an initial (starting) mosaic with $\bar{v}^*_I(M) < 4$ then the cell corner degree $\bar{v}^*$ increases strictly monotonically over time.*

**Proof:**

Since there are two steps in the model ($R_0$ and $R_1$), so, if we can show separately for both that the cell corner degree is increases in the case of an initial condition *with $\bar{v}_I^*(M) < 4$*, then we have confirmed the statement of Lemma 3.

In the case of $R_0$-type steps we have $N_{k_0}^* = N_{k_0-1}^* + 4;\ F_{k_1} = F_{k_1-1} + 1;\ V_{k_1} = V_{k_1-1} + 2$ (see Table 1). Based on formula (4) we see that if in the initial state $\bar{v}_I^*(M) < 4$, $\bar{v}^*$ will increase in the first step. Also, we have $\bar{v}^* < 4$ after any finite amount of steps and, as a consequence, $\bar{v}^*$ is increasing strictly monotonically under $R_0$ type steps.

In the case of $R_1$-type steps we have $N_{k_1}^* = N_{k_1}^* + 1;\ F_{k_2} = F_{k_2-1};\ V_{k_2} = V_{k_2-1}$ (see Table 1), so, based on formula (4), here the cell corner degree $\bar{v}^*$ is always increasing, regardless of the initial value.

Q.e.d.

## 4. Interpretation of the results potential applications

Lemma 1 establishes the $\bar{v}^* = 2\bar{n}^*$ line as the global attractor for $p$-trajectories, i.e. for fracture pattern evolution processes where the relative weight of secondary fracture and crack healing/rearrangement remains constant over time. While physical evolution processes may be much more complex, it appears to be feasible to approximate them piecewise by $p$-trajectories. All such general processes will ultimately converge onto a point of the $\bar{v}^* = 2\bar{n}^*$ line and this suggests that this line may have indeed significance, even beyond $p$-trajectories. This observation is underlined by Lemma 2 which states that cell density is growing monotonically over time. The $\bar{v}^* = 2\bar{n}^*$ line represents convex mosiacs with maximal cell density $\bar{\rho}(M) = 2$.

Since cell density evolves monotonically, it defines a hierarchy among fracture patterns. This hierarchy may be defined even more sharply by connecting mosaics with a $p$-trajectory; in this respect we formulate

**Conjecture 1:**

*Let $M_1$ and $M_2$ be two infinite, convex, balanced mosaics. Then there exists a unique value $0 \leq p(M_1, M_2) \leq 1$ such that p-trajectory connects $M_1$ and $M_2$.*

In this case we can say that $M_1$ and $M_2$ are *p*-related. While these considerations are neither entirely rigorous nor are they supported by experiments, our model and the results derived from this model suggest that the *maturity* of a fracture network may be related to the cell density $\bar{\rho}(M)$ of the convex mosaic representing it.

In Figure 4 we show two mud crack patterns (discussed in detail in [2], shown in Figure 1 (a9 and (b) of the current article) connected by a *p=0.685* trajectory in the (b)→(a) direction. While this connection certainly does not indicate that pattern (a) evolved from pattern (b), it does tell us that the inverse would not be compatible with the model and this might help a meaningful geophysical comparison.

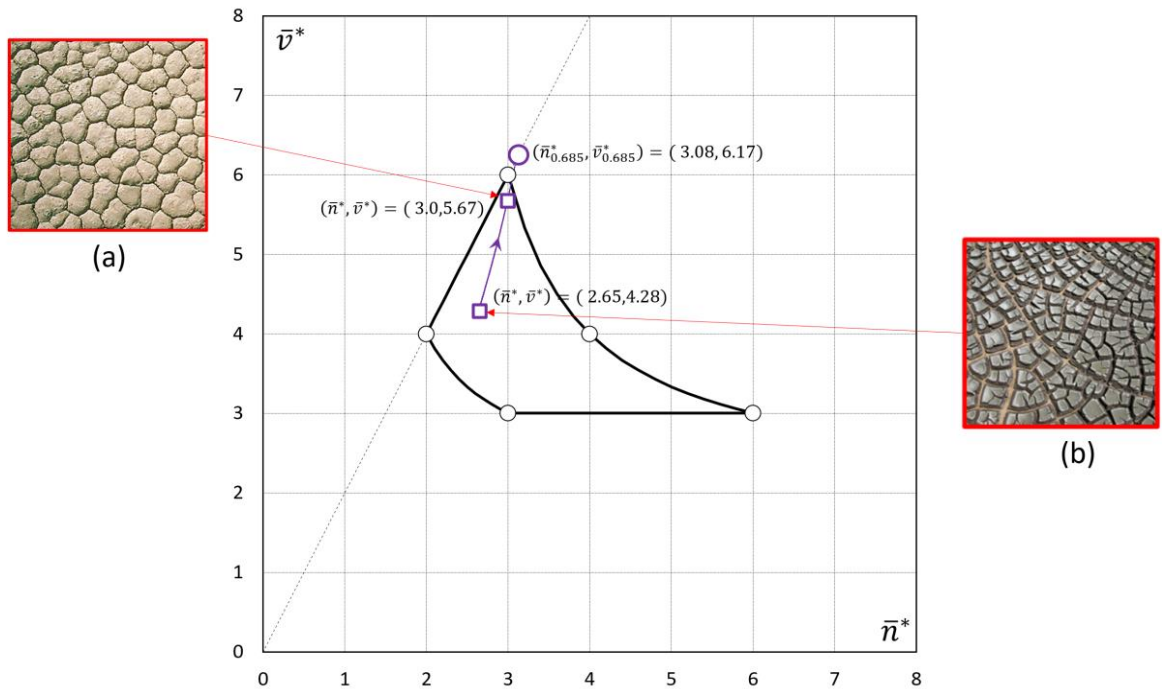

**Figure 4:** Mud crack patterns shown in Figure 1(a) and (b) of this paper (from Figure 3 in Article [2]), connected by a *p=0.685* trajectory in the direction (b)→(a).

## **Acknowledgement**

The authors thank Marjorie Senechal for the discussion and helpful comments. The support of the NKFIH Hungarian Research Fund grant 134199 and of the NKFIH Fund TKP2021 BME-NVA, carried out at the Budapest University of Technology and Economics, is kindly acknowledged. Krisztina Regös: This research has been supported by the program ÚNKP-22-3 by ITM and NKFIH. The gift representing the Albrecht Science Fellowship is gratefully appreciated.